\documentclass[,final] {aipproc}

\layoutstyle{6x9}

\usepackage{times,fancyhdr,bbold}
\usepackage{graphicx}
\usepackage{subfig}

\begin{document}

\newcommand{\dif}{{\rm{d}}}
\newcommand{\e}{\rm{e}}
\newcommand{\B}{\mathcal{B}}
\newcommand{\R}{\mathcal{R}}
\newcommand{\be}{\begin{equation}}
\newcommand{\ee}{\end{equation}}
\newcommand{\bi}{\begin{itemize}}
\newcommand{\ei}{\end{itemize}}
\newcommand{\med}{\medskip \\}
\newcommand{\bq}{\mathbf{q}}
\newcommand{\bk}{\mathbf{k}}
\newcommand{\bl}{\mathbf{l}}
\newcommand{\non}{\nonumber\\}
\newcommand{\asbar}{\bar{\alpha}_s}
\newcommand{\K}{\mathcal{K}}

\title{The hard to soft Pomeron transition in small $x$ DIS data using optimal renormalization}

\classification{}
\keywords      {}

\author{Clara Salas}{
  address={Instituto de F{\' \i}sica Te{\' o}rica UAM/CSIC 
and   Universidad Aut{\' o}noma de Madrid, \\  \normalsize   
E-28049 Madrid, Spain}
}

\begin{abstract}
   We show that it is possible to describe the effective Pomeron
  intercept, determined from the HERA Deep Inelastic Scattering data
  at small values of Bjorken $x$, using next-to-leading order BFKL
  evolution together with collinear improvements.  To obtain a good
  description over the whole range of $Q^2$ we use a non-Abelian
  physical renormalization scheme with BLM optimal scale, combined
  with a parametrization of the running coupling in the infrared
  region.
\end{abstract}

\maketitle


\section{Introduction}

The description of Deep Inelastic Scattering (DIS) data for the
structure function $F_2 (x, Q^2)$ in different regions of Bjorken $x$ and
virtuality of the photon $Q^2$ is one of the classical problems in
perturbative QCD. The recent HERA combined results~\cite{Aaron:2009aa} for $F_2$ cover a broad range of values of the photon virtuality $Q^2$ and also reaches very small values of $x$, making this observable suitable to study the high energy limit, given when the center of mass energy of the system is much higher than any other scale involved or, in other words, the region of low Bjorken $x$.  

The aim of the present study is to analyze the structure function $F_2$ and Pomeron intercept within perturbative QCD in this mentioned region of low values of $x$ using high energy factorization~\cite{Catani:1990eg}, which convolutes the proton and photon impact factors with the gluon Green's function. For the latter we use the solution to the BFKL evolution equation at Next-to-Leading-Logarithmic (NLL) accuracy~\cite{Fadin:1998py,Ciafaloni:1998gs} and introduce collinear improvements, needed to deal with the leading collinear singularities that are numerically large in this kinematic region~\cite{Salam:1998tj,Ciafaloni:1999yw,Ciafaloni:2003rd, Altarelli:2005ni,Vera:2005jt}. We show how the use of a non-Abelian physical renormalization scheme with optimal scale setting\footnote{We use in this work MOM scheme with BLM optimal scale~\cite{Brodsky:1982gc,Brodsky:1998kn} but there are other similar choices.} allows for a good description of the Pomeron over the full range of $Q^2$.

A numerical analysis of the Pomeron intercept $\lambda$ and comparison with data is given, using the prediction of Regge theory that claims that in the high energy limit the asymptotic expression for $F_2$ should grow with energy as $(s/s_0)^{\lambda}=(1/x)^{\lambda}$ . This analysis is presented in more detail in~\cite{Hentschinski:2012kr}.

\section{Theoretical setup}

\begin{figure}
\hspace{-1cm}
 \includegraphics[width=.25\textwidth]{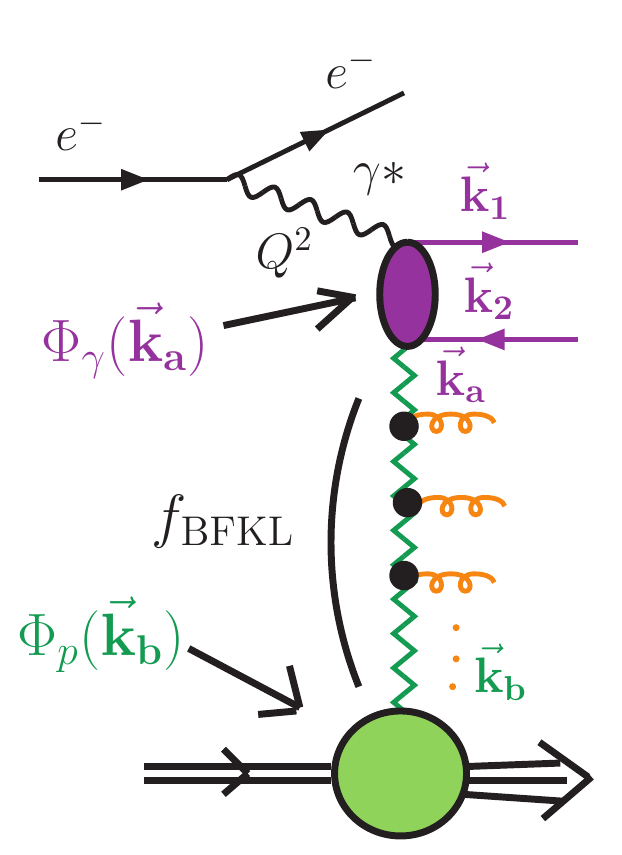} \hspace{.5cm}
  \includegraphics[width=.4\textwidth]{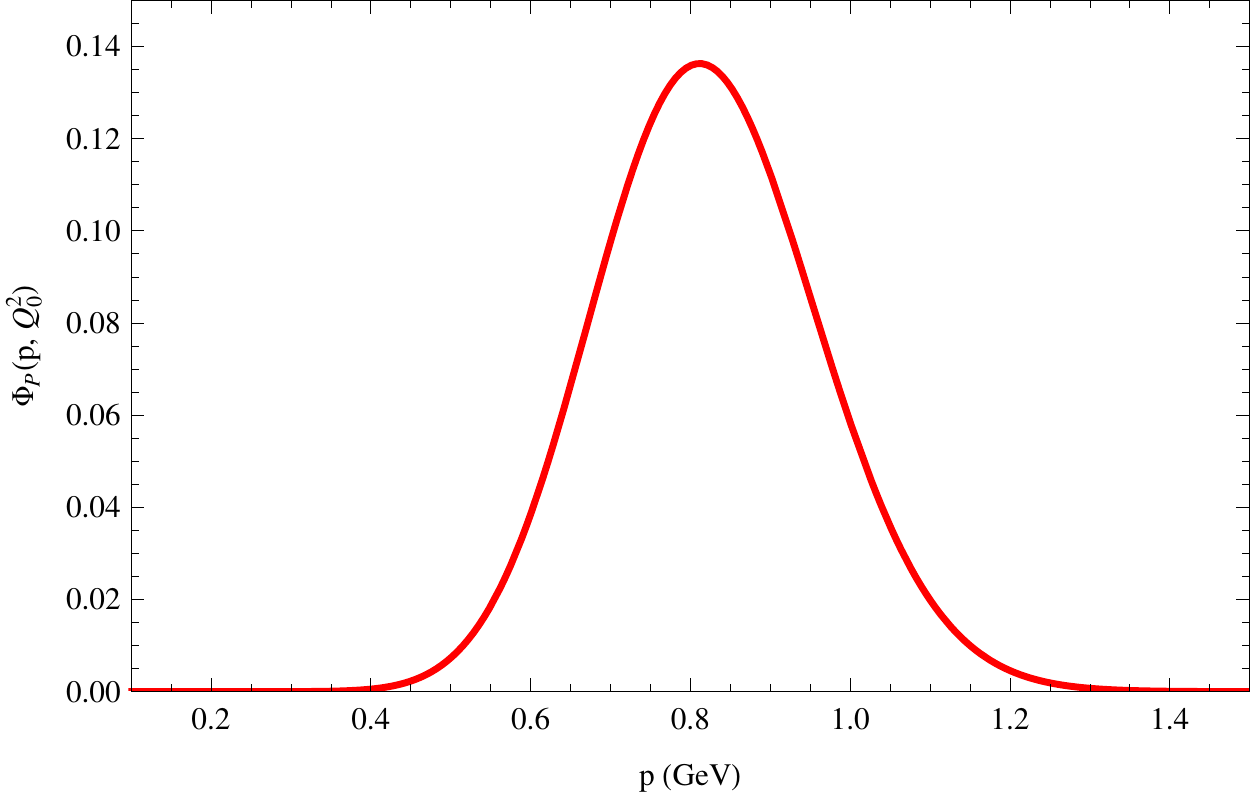}\hspace{.5cm}
  \includegraphics[width=.4\textwidth]{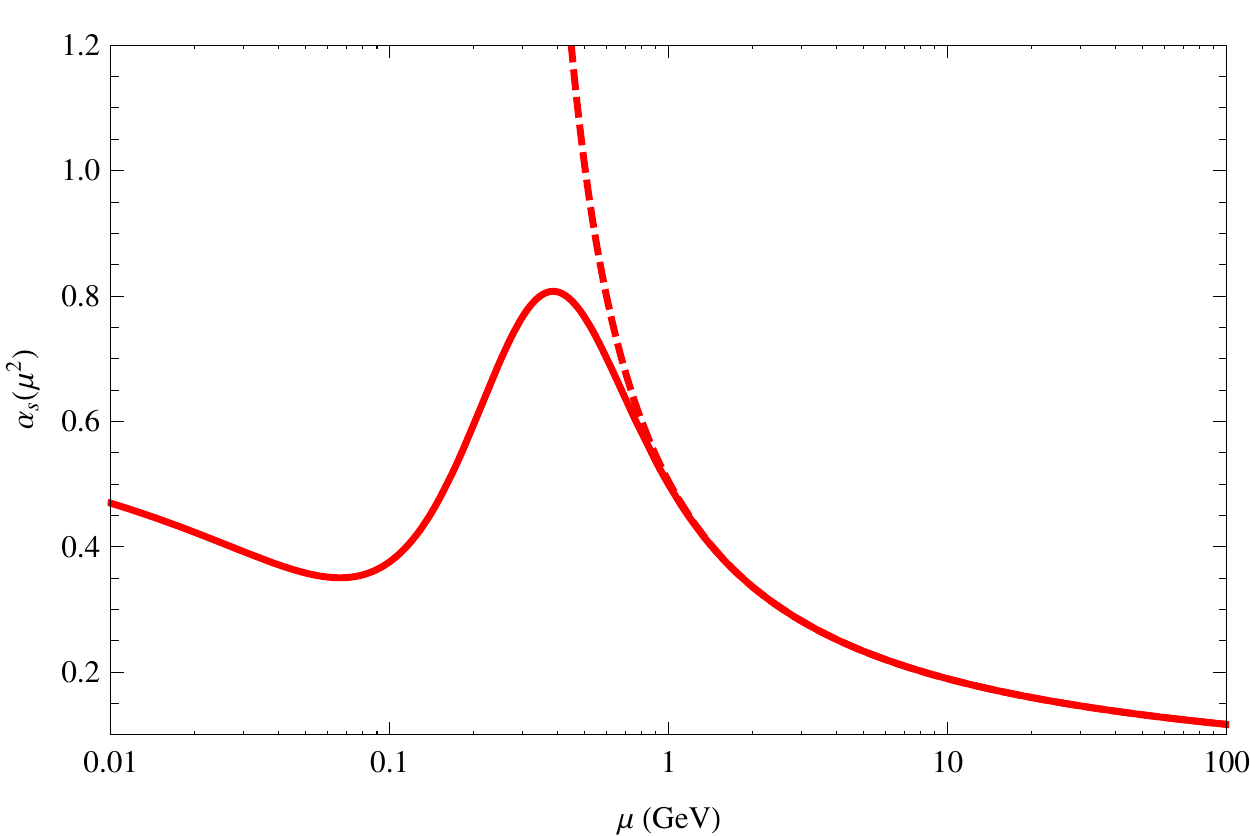}
 \caption{High energy factorization (left $\equiv$ a), proton impact factor (center $\equiv$ b) and model for the running coupling freezing in the IR (right $\equiv$ c).}
  \label{fig:factorization}
\end{figure}

High energy factorization allows to write the proton structure function as a convolution in transverse momentum space of a non perturbative object describing the proton (proton impact factor $\Phi_p$) with the photon (photon 
impact factor $\Phi_\gamma$), calculated using perturbation theory, together with a gluon Green's function $f$, 
linking both process-dependent components and accounting for the BFKL evolution, as shown in fig.~\ref{fig:factorization}a:
{\small
$$
F_2(x,Q^2)=\frac{F_c}{(2\pi)^4}\int\frac{\dif^2\bk_a}{\bk^2_a}\int\frac{\dif^2\bk^2_b}{\bk_b}\Phi_{\gamma}(\bk_a)\, f(x,\bk_a,\bk_b)\,\Phi_{p}(\bk_b) \; . 
$$
}
We use for simplicity in our analysis the leading order photon impact factor as presented in~\cite{Forshaw:1997dc}. Our choice for the proton impact factor is a Poissonian-like distribution given by (see Fig.~\ref{fig:factorization}b)
{\small
$$
\Phi_P(\bk)=A_p (\bk^2/Q_0^2)^\delta \e^{-\bk^2/Q_0^2}
$$ 
}
Finally, the gluon Green's function is governed by the BFKL equation. Since in DIS $Q^2 \gg Q_0^2$, it has to be written in a form consistent with the resummation of $\bar{\alpha}_s \log{(1/x)}$ contributions:
{\small
$$
f (s, q , p) = \frac{1}{2 \pi q^2} \int \frac{d \omega}{2 \pi i} \int \frac{d \gamma}{2 \pi i} 
 \left(\frac{{q^2}}{p^2}\right)^{\gamma} \left(\frac{s}{q^2}\right)^\omega \frac{1}{\omega- K\left(\gamma - \frac{\omega}{2}\right)},
$$
}
where $K$ is the NNL BFKL kernel. The zeros of the denominator in the integrand generate in the limits $\gamma \to 0,1$ all-orders terms not compatible with DGLAP evolution~\cite{Salam:1998tj,Vera:2005jt}. The first of these pieces (${\cal O} (\alpha_s^2)$) is removed when the NLO correction to the BFKL kernel is taken into account but not the 
higher order ones, which remain and are numerically important. A scheme to eliminate these spurious contributions~\cite{Salam:1998tj} consists of modifying the BFKL kernel by making a shift of the form $K(\gamma) \to K(\gamma+\omega/2)$. 

Once we include NLL corrections with the collinear improvement as described in~\cite{Vera:2005jt} we find the following expression for $F_2$ in Mellin space:
{\small
\begin{eqnarray}
F_2 (x,Q^2) &\propto & \int_{-\infty}^\infty \dif \nu \,  x^{-{\chi} \left(\frac{1}{2} + i\nu \right)} c_2 (\nu)\; c_P (\nu) \Bigg\{1 + \bar{\alpha}_s^2 \log{\left(\frac{1}{x}\right)} \frac{\beta_0}{8 N_c} \chi_0 \left(\frac{1}{2} + i \nu\right) 
\\
&& \hspace{-1.5cm}
\Bigg[-\log{\left(\frac{Q^2 Q_0^2}{\mu^4}\right)}- \psi \left(\delta-\frac{1}{2}-i \nu\right) +  i \Bigg(\pi {\rm coth} (\pi \nu)-2 \pi \tanh{(\pi \nu)}- M_2 (\nu)\Bigg)\Bigg]\Bigg\}\; , \nonumber
\label{Frho}
\end{eqnarray}
}
where $c_2(\nu)$ and $c_P(\nu)$ are the photon and proton impact factors in $\nu$-space, respectively and $M_2$ is a simple function of $\nu$ defined in~\cite{Hentschinski:2012kr}. Note that in this equation we have exponentiated only the scale invariant terms of the kernel. The terms that break the invariance are due to the running of the coupling and appear as a differential operator in $\nu$ acting on the impact factors~\cite{Vera:2007dr}.

We introduce the running of the coupling in a way that removes the $\mu$ dependence of eq.~\eqref{Frho} by making the replacement
{\small
$$
\bar{\alpha}_s - \bar{\alpha}_s^2 \frac{\beta_0}{8 N_c} \log{\left(\frac{Q^2 Q_0^2}{\mu^4}\right)} \,\,\, \to \,\,\, 
\bar{\alpha}_s \left(Q Q_0\right)\; .
$$
}
This resummed coupling is consistent with the Landau one up to NLO accuracy. We see in our analysis that in order to have a good description of the Pomeron intercept over the full range $1\;{\rm GeV}^2 < Q^2 < 300 \;{\rm GeV}^2$, we need to move to a renormalization scheme based on the existence of a possible IR fixed point. The pieces of the NLL BFKL kernel proportional to $\beta_0$ are absorbed in a new definition of the running coupling so that all the vacuum polarization effects  from the $\beta_0$ function are resummed. We refer to the reader to~\cite{Brodsky:1982gc,Brodsky:1998kn} for details. 

As a final ingredient, we use a simple parametrization of the running coupling introduced by Webber in~\cite{Webber:1998um} which is consistent with global data of power corrections to perturbative observables and with the usual running coupling with Landau pole up to NLO accuracy. Fig.~\ref{fig:factorization}c shows both models, being the solid curve the new one.

\section{Numerical analysis \& comparison to DIS data}

\begin{figure}
 \hspace{-1.5cm}
 \includegraphics[width=.65\textwidth]{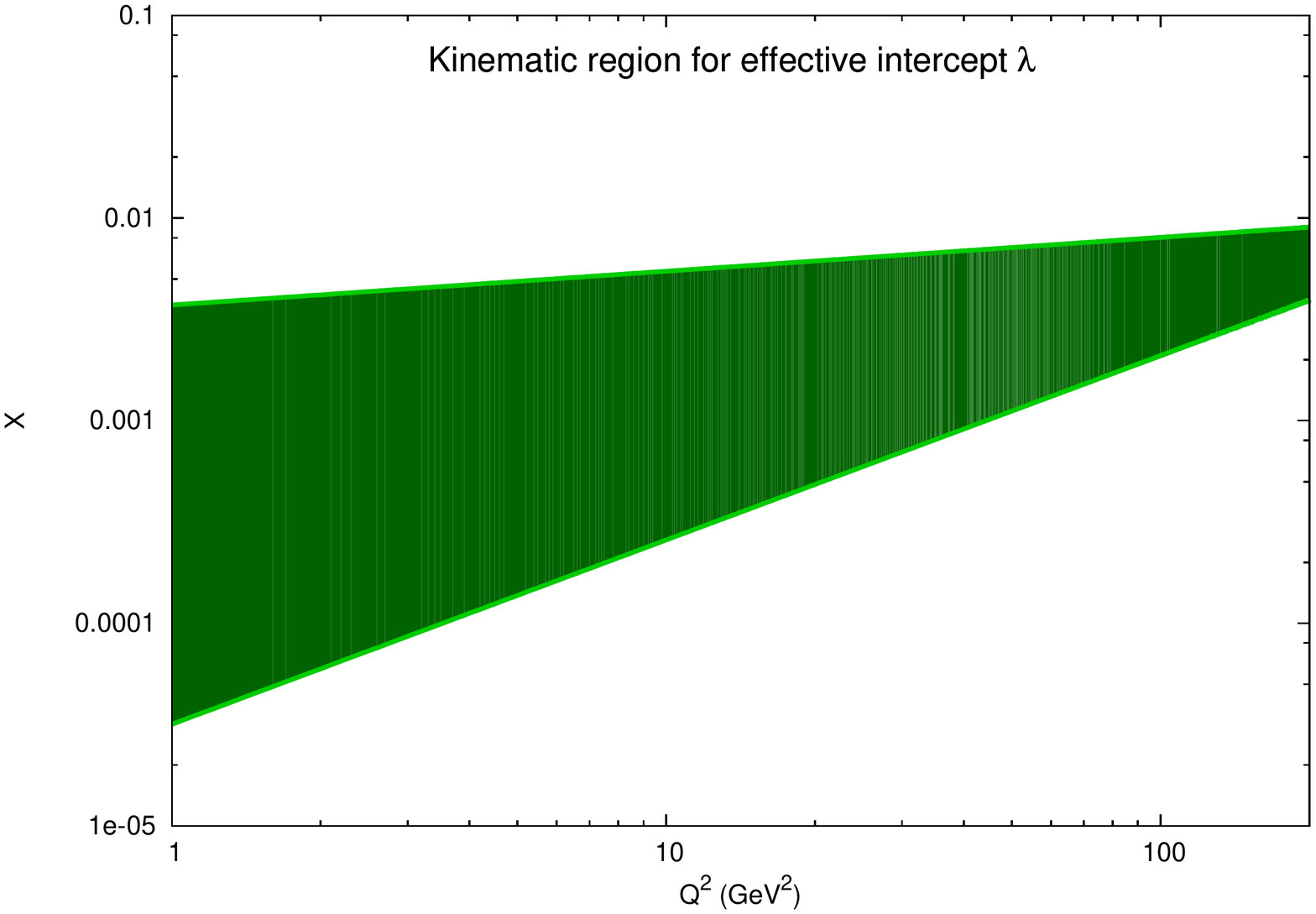} \hspace{-1cm}
  \includegraphics[width=.65\textwidth]{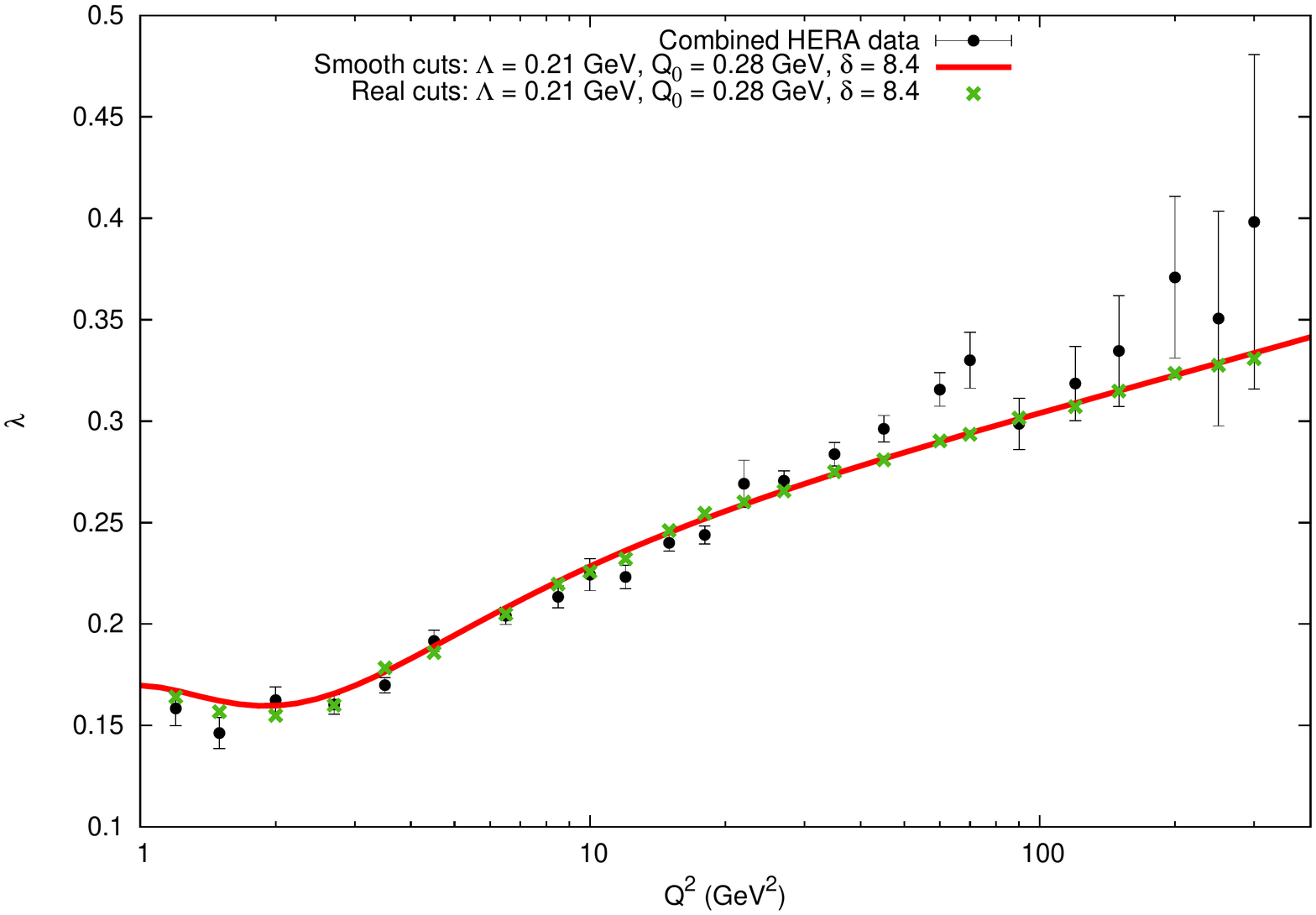}
 \caption{Experimental bounds in $x$ (left $\equiv$ a) and evolution of $\lambda$ with $Q^2$ (right $\equiv$ b).}
  \label{Domain}
\end{figure}

To obtain our theoretical results we have calculated the logarithmic
derivative $\frac{d \log F_2 }{ d \log (1/x)}$ using Eq.~(\ref{Frho}) with some
modifications.  For the comparison
with DIS data we chose the
values $Q_0 = 0.28 \,{\rm GeV}$ and $\delta = 8.4$, $n_f=4$ and $\Lambda= 0.21\, {\rm GeV}$ (see fig.~\ref{Domain}b). The experimental input has been derived from the
combined analysis performed by H1 and ZEUS in Ref.~\cite{Aaron:2009aa}
with $x<10^{-2}$. In the results indicated with ``Real cuts'' we have
calculated the effective intercept for $F_2$ at a fixed $Q^2$,
averaging its values in a sample of $x$ space consistent with the actual
experimental  cuts in $x$. To generate the continuous line with
label ``Smooth cut'' we have used as boundaries in $x$ space those
shown in Fig.~\ref{Domain}a, which correspond to an interpolation of
the real experimental boundaries. Note that the difference between
both approaches is very small. We would like to stress the accurate description of the combined HERA data in our approach, in particular at very low values of $Q^2$.

It is possible to improve the quality of our
fit by introducing subleading contributions such as threshold effects
in the running of the coupling, heavy quark masses and higher order
corrections to the photon impact factor which became recently
available \cite{Balitsky:2012bs}. We are presently working on these improvements, together with a
comparison to data not averaged over $x$, also including an analysis of $F_L$.


\begin{theacknowledgments}

I acknowledge the European Comission under contract LHCPhenoNet (PITN-GA-2010-264564), the Comunidad de Madrid through Proyecto HEPHACOS ESP-1473, and MICINN (FPA2010-17747).

\end{theacknowledgments}

\bibliographystyle{aipproc}  

\bibliography{citations}

\IfFileExists{\jobname.bbl}{}
 {\typeout{}
  \typeout{******************************************}
  \typeout{** Please run "bibtex \jobname" to optain}
  \typeout{** the bibliography and then re-run LaTeX}
  \typeout{** twice to fix the references!}
  \typeout{******************************************}
  \typeout{}
 }

\end{document}